# How to explore replicator equations?

G.P. Karev

Lockheed Martin MSD, National Institute of Health,
Bldg. 38A, Rm. 5N511N, 8600 Rockville Pike, Bethesda, MD 20894, USA

E-mail: *karev@ncbi.nlm.nih.gov*

**Abstract.** Replicator equations (RE) are among the basic tools in mathematical theory of selection and evolution. We develop a method for reducing a wide class of the RE, which in general are systems of differential equations in Banach space to "escort systems" of ODEs that in many cases can be explored analytically. The method has potential for different applications; some examples are given.

## 1. Introduction

Consider a system formed by varieties, each of which is characterized by the specific value of the vector-parameter $\mathbf{a} = (a_1,...a_n)$. In general, vector $\mathbf{a}$ can be considered as a microstate of the system; the parameters $a_i$ may have different origin. Let $l(t,\mathbf{a})$ be the density of individuals in the state $\mathbf{a}$ at the moment $t$, so that $\int_v l(t,\mathbf{a})d\mathbf{a}$ is the number of individuals with values of $\mathbf{a}$ in the phase volume $v$, and $N(t) = \int_A l(t,\mathbf{a})d\mathbf{a}$ is the total population size at $t$ moment. The dynamics of the system is defined by the following equations

$$dl(t,\mathbf{a})/dt = l(t,\mathbf{a})F(t,\mathbf{a}), \qquad (1.1)$$

$$P_t(\mathbf{a}) = l(t,\mathbf{a})/N(t).$$

The reproduction rate ("fitness") $F(t,\mathbf{a})$ is supposed to be a smooth function of $t$ and a measurable function of $\mathbf{a}$; it may depend on some "extensive variables", which are some averages over $P_t(\mathbf{a})$. The initial distribution $P_0(\mathbf{a})$ and the initial population size $N(0)$ are supposed to be given.

It is known that $N(t)$ satisfies the equation $dN/dt = NE^t[F]$; here and below we use the notation $E^t[\varphi] = \int_A \varphi(\mathbf{a})P_t(\mathbf{a})d\mathbf{a}$. It is also known [10] that $P_t(\mathbf{a})$ solves the *replicator equation*

$$dP_t(\mathbf{a})/dt = P_t(\mathbf{a})(F(t,\mathbf{a}) - E^t[F(t,.)]) \qquad (1.2)$$

and that the solution of RE at given initial distribution $P_0(\mathbf{a})$ is unique (if it exists).

In the last decades it was discovered that replicator equations appear not only in population genetics and selection theory [2], but also in very different areas, such as theoretical ecology [9], dynamical game theory [4] and even in some physical problems, see the survey [3]. Most of these applications assume that the fitness depends linearly on the frequencies. Here we show that a wide class of replicator equations including those with the "linear fitness" can be solved explicitly, and the solution has a form of time-dependent Boltzmann distribution. The obtained results are applied to some particular selection systems and corresponding replicator equations.

## 2. The method

If the reproduction rate $F(t,\mathbf{a})$ is known explicitly as a function of $t$, then the RE can be easily solved:

$$P_t(\mathbf{a}) = \frac{\exp(\Phi(t,\mathbf{a}))}{Z(t)} P_0(\mathbf{a})$$

where $\Phi(t,\mathbf{a}) = \int_0^t F(u,\mathbf{a})du$ and $Z(t) = E^0[\exp(\Phi(t,.))]$.

Generally, the reproduction rate $F(t,\mathbf{a})$ is not given as an explicit function and should be computed depending on the current population characteristics. For example, widely used logistic models have the reproduction rate of the form $F(t,\mathbf{a}) = \varphi(\mathbf{a})(1 - N(t)/B)$ where $B$ is the upper boundary of the population size. So we should explore the selection systems with the reproduction rate that can depend on some integral characteristics of the system. We account for extensive characteristics in the form

$$G_i(t) = \int_A g_i(\mathbf{a})l(t,\mathbf{a})d\mathbf{a} = N(t)E^t[g_i], \tag{2.1}$$

which depend on the total system size and population densities, and intensive characteristics in the form

$$H_k(t) = \int_A h_k(\mathbf{a})P(t,\mathbf{a})d\mathbf{a} = E^t[h_k], \tag{2.2}$$

which do not depend on the system size but only on the population frequencies. We will refer to both of them as to "regulators" for brevity. Finally, we have the following general version of the master model:

$$dl(t,\mathbf{a})/dt = l(t,\mathbf{a})F(t,\mathbf{a}), \tag{2.3}$$

$$F(t,\mathbf{a}) = \sum_{i=1}^{n} u_i(t,G_i)\varphi_i(\mathbf{a}) + \sum_{k=1}^{m} v_k(t,H_k)\psi_k(\mathbf{a}),$$

$$P(t,\mathbf{a}) = l(t,\mathbf{a})/N(t)$$

where $u_i, v_k$ are appropriate functions. The initial pdf $P(0,\mathbf{a})$ and the population size $N(0)$ need be given.

The system distribution $P(t,\mathbf{a})$ solves the replicator equation

$$dP_t(\mathbf{a})/dt = P_t(\mathbf{a})(F(t,\mathbf{a}) - E^t[F(t,.)]) \tag{2.4}$$

where now $E^t[F(t,.)] = \sum_{i=1}^{n} u_i(t,G_i)E^t[\varphi_i] + \sum_{k=1}^{m} v_k(t,H_k)E^t[\psi_k]$ and all regulators $G_i$, $H_k$ together with $E^t[\varphi_i]$, $E^t[\psi_k]$ are not given functions of time and should be determined.

Model (2.3) was studied in [8] (see also [7] for discrete time version). The developed theory yields an effective algorithm for investigation of selection systems within frameworks of model (2.3) and for solving of replicator equation (2.4). Let us describe the main steps of the algorithm.

Consider the probability space $\{A, \mathbf{A}, P(0,\mathbf{a})\}$ and define the functional

$$M(r;\boldsymbol{\lambda},\boldsymbol{\delta}) = \int_A r(\mathbf{a})\exp(\sum_{i=1}^n \lambda_i\varphi_i(\mathbf{a}) + \sum_{k=1}^m \delta_k\psi_k(\mathbf{a}))P(0,\mathbf{a})d\mathbf{a} \qquad (2.5)$$

for measurable functions $r$ on the space $\{A,\mathbf{A},P(0,\mathbf{a})\}$; all the functions $\varphi_i(\mathbf{a}),\psi_k(\mathbf{a})$ are also supposed to be measurable on this space.

Define the auxiliary variables $q_i, s_k$ by the "escort system" of ODE

$$dq_i/dt = u_i(t, G_i^*(t)), q_i(0) = 0, i = 1,\ldots n, \qquad (2.6)$$

$$ds_k/dt = v_k(t, H_k^*(t)), s_k(0) = 0, k = 1,\ldots m$$

where we denote $G_i^*(t) = N(0)M(g_i;\mathbf{q}(t),\mathbf{s}(t))$, $H_k^*(t) = M(h_k;\mathbf{q}(t),\mathbf{s}(t))/M(1;\mathbf{q}(t),\mathbf{s}(t))$.

Let $K_t(\mathbf{a}) = \exp(\sum_{i=1}^n q_i(t)\varphi_i(\mathbf{a}) + \sum_{k=1}^m s_k(t)\psi_k(\mathbf{a}))$; then the solution to system (2.3)

$$l(t,\mathbf{a}) = l(0,\mathbf{a})K_t(\mathbf{a});$$

$$G_i(t) = G_i^*(t), H_k(t) = H_k^*(t);$$

$$N(t) = N(0)M(1;\mathbf{q}(t),\mathbf{s}(t));$$

$$P(t,\mathbf{a}) = P(0,\mathbf{a})K_t(\mathbf{a})/E^0[K_t]. \qquad (2.7)$$

Formula (2.7), which gives the solution of replicator equation (2.4) is the central result of the theory.

The general method is simplified in an important case of the reproduction rate $F(t,\mathbf{a}) = \sum_{i=1}^n f_i(t,S_i)\phi_i(\mathbf{a})$ with the regulators $S$ of the forms $N(t), E^t[\phi_i], N(t)E^t[\phi_i]$ only. In this case we can use the moment generating function of the joint initial distribution of the variables $\{\phi_i\}$ only, $M_0(\boldsymbol{\lambda}) = E^0[\exp(\sum_{i=1}^n \lambda_i\phi_i)]$ instead of general functional (2.5). The escort system reads

$$dq_i/dt = f_i(t, S_i(t)), q_i(0) = 0, i = 1,\ldots n$$

where $S_i(t)$ are defined with the help of formulas $N(t) = N(0)M_0(\mathbf{q}(t))$, $E^t[\varphi_k] = \partial_k \ln M_0(\mathbf{q}(t))$. Here we denoted $\partial_k M_0(\boldsymbol{\delta}) = \partial M_0(\boldsymbol{\delta})/\partial \delta_k$ for brevity. The solution of corresponding replicator equation

$$P(t,\mathbf{a}) = P(0,\mathbf{a})K_t(\mathbf{a})/E^0[K_t], \quad E^0[K_t] = M_0(\mathbf{q}(t)).$$

The following examples demonstrate the algorithm at work.

## 3. Applications and examples

1. ***Inhomogeneous Malthusian model and the model of early evolution***

The simplest replicator equation with a single continuous parameter reads

$$dP_t(a)/dt = P_t(a)(a - E^t[a]).$$

The corresponding selection system is the inhomogeneous Malthusian model

$$dl(t,a)/dt = al(t,a).$$

Let $M(\lambda) = \int_A \exp(\lambda a) P_0(a) da$. Then the solution of the model $l(t,a) = \exp(at)l(0,a)$, $N(t) = N_0 M(t)$, and the solution to the replicator equation $P_t(a) = P_0(a)\exp(at)/M(t)$.

Solutions of inhomogeneous Malthusian and logistic models and their applications were studied in [5, 6, 8]. It was shown that even in these simplest cases the replicator equations possess a variety of solutions depending on the initial distribution, which may have many interesting and even counterintuitive peculiarities. Let us demonstrate some of them on the following inhomogeneous Malthusian model with limiting factors.

A model of early biological evolution was suggested in [12]. Each organism is characterized by the vector **a** where the component $a_i$ is the thermodynamic probability that protein $i$ is in its native conformation. In order to study the connection between molecular evolution and population, the authors supposed that organism death rate $d$ depends on the stability of its proteins as $d = d_0(1 - \min a_i)$, $d_0$=const. Hence, neglecting possible mutations (accounted for by the authors in their simulations), the model can be formalized as the system

$$dl(t,\mathbf{a})/dt = l(t,\mathbf{a})B(m(\mathbf{a}) - a_0)), \qquad (3.1)$$

where $m(\mathbf{a}) = \min[a_1,...a_n]$, $B = b/(1-a_0)$, $b$ is the birth rate, $a_0$ is the native state probability of a protein. In what follows we put $B = 1$ for simplicity. Following [12] suppose that the values $a_i$ are independent of each other; we can consider $a_i$ as the $i$-th realization of a random variable with common pdf $f(a)$. Then, it is well known that the pdf of $m = \min[a_1,...a_n]$ where $a_i$ are independent identically distributed random variables is $g(m) = n(1-G(m))^{n-1}f(m)$ where $G(m) = \int_0^m f(a)da$ is the cumulative distribution function. The equation $dl(t,m)/dt = l(t,m)(m - a_0)$ is a version of inhomogeneous Malthusian equation and can be solved explicitly. In particular, if

$$f(a) = \exp(-a/T)/Z, \quad Z = \sum_a \exp(-a/T) \qquad (3.2)$$

is the Boltzmann distribution with $a > 0$, then

$$g(a) = n(1-G(a))^{n-1}f(a) = (n(\sum_{x>a}\exp(-x/T))^{n-1})\exp(-a/T)/\sum_x\exp(-x/T)). \qquad (3.3)$$

For distribution (3.2) with continuous range of values of $a$, $a \in (0,\infty)$, $Z = T$, $1 - G(m) = \exp(-m/T)$ and

$$g(m) = n(\exp(-m(n-1)/T)\exp(-m/T)/T = n/T \exp(-mn/T). \qquad (3.4)$$

If $a \in (0, E)$, then $Z = T(1 - \exp(-E/T))$, $1 - G(m) = \dfrac{\exp((E-m)/T) - 1}{\exp(E/T) - 1}$, and

$$g(m) = \frac{n\exp(-m/T)}{T(1 - \exp(-E/T))}[\frac{1 - \exp((E-m)/T)}{1 - \exp(E/T)}]^{n-1}. \qquad (3.5)$$

Let $M_0(\lambda) = E^0[\exp(\lambda m)]$. For initial distribution (3.4), $M_0(\lambda) = \dfrac{1}{1 - \lambda T/n}$. Hence,

$$l(t,m) = l(0,m)\exp((m - a_0)t),$$

$$N(t) = N(0)\exp(-a_0 t)\frac{1}{1 - tT/n},$$

$$P(t,\mathbf{a}) = P(0,\mathbf{a})(1 - tT/n)\exp(m(\mathbf{a})t).$$

At the moment $t_{max} = n/T$ the population „blows up": $N(t)$ and $l(t,\mathbf{a})$ tend to infinity at $t \to t_{max}$. Let us denote $p(t,a) = P(t,\{\mathbf{a}: m(\mathbf{a}) = a\})$. Then at $t < t_{max}$

$$p(t,a) = n/T \exp(-an/T + at)(1 - tT/n) = (n/T - t)\exp(-a(n/T - t)).$$

The probability $P(t,\{\mathbf{a}: m(\mathbf{a}) < a\})$ tends to 0 for any finite $a$ at $t \to t_{max}$. Loosely speaking, the total "probability mass" goes to infinity after a finite time interval. So, we should conclude that model (3.1), (3.2) which allows arbitrary large values of the parameter $a$ with nonzero probability has no "physical" sense.

This problem can be eliminated by taking the initial distribution (3.5), which allows only bounded values of the parameter $a$. For pdf (3.5), the integral $M_0(\lambda) = \int_0^E \exp(\lambda x) g(x) dx$ is well defined for any $\lambda$ but not expressed in quadratures. Nevertheless we can obtain much information about the system distribution and its dynamics. The current pdf

$$p(t,a) = \frac{n}{T(\exp(E/T) - 1)^n} \exp((E-a)/T)(\exp((E-a)/T) - 1)^{n-1} \frac{\exp(at)}{M_0(t)}$$

where $M_0(t)$ is finite for all $t$. So, the pdf is well determined at any time moment, in contrast to the previous case. The total distribution concentrates with time at the point $a = E$, which provides the maximal reproduction rate.

2. *The Fisher-Haldane-Wright equation*

It seems that one of the first replicator equations was introduced by R. Fisher in 1930 [1] for genotype evolution:

$$\frac{dp_a}{dt} = p_a(W_a - W), \quad a = 1,...n \tag{3.6}$$

where $W_a = \sum_b W_{ab} p_b$, $W = \sum_{a,b} p_a W_{ab} p_b$. Here $p_a$ is the frequency of the gamete $a$, $W_{ab}$ is the absolute fitness of the zygote $ab$. In mathematical genetics this equation is known as the Fisher-Haldane-Wright equation (FHWe) and sometimes is referred to as the main equation of mathematical genetics (see, [11]).

The matrix $\{W_{ab}\}$ is symmetric and hence has the spectral representation

$$W_{ab} = \sum_{k=1}^m \omega_k h_k(a) h_k(b) \tag{3.7}$$

where $\omega_k$ are non-zero eigenvalues and $h_k$ are corresponding orthonormal eigenvectors of $\mathbf{W}$; $m$ is the rank of $\{W_{ab}\}$. Then

$$W_a(t) = \sum_{b=1}^n W_{ab} p_b(t) = \sum_{k=1}^m \sum_{b=1}^n \omega_k h_k(a) h_k(b) p_b(t) = \sum_{k=1}^m \omega_k E^t[h_k] h_k(a).$$

The FHW-equation now reads

$$\frac{dp_a}{dt} = p_a(\sum_{k=1}^m \omega_k (h_k(a) E^t[h_k] - (E^t[h_k])^2), \quad a = 1,...n.$$

Consider the associated selection system:

$$dl(t,a)/dt = l(t,a)\sum_{k=1}^{m}\omega_k h_k(a)E^t[h_k] \qquad (3.8)$$

The range of values of the parameter $a$ is now a finite set, unlike in the previous examples. Define the mgf of the initial distribution of the parameter $a$:

$$M_0(\boldsymbol{\delta}) = \sum_{i=1}^{n} \exp(\sum_{k=1}^{m}\delta_k h_k(a))P(0;a) = E^0[\exp(\sum_{k=1}^{m}\delta_k h_k(.))].$$

Compose and solve the escort system of ODE

$$ds_i/dt = \omega_i E^0[h_i(.)\exp(\sum_{k=1}^{m}s_k h_k(.))]/E^0[\exp(\sum_{k=1}^{m}s_k h_k(.))],\ i=1,...m.$$

These equations can be written in a more compact form
$$ds_i/dt = \omega_i \partial \ln M_0(\mathbf{s})/\partial s_i.$$

Then the solution to the selection system (3.8)

$$l(t,a) = l(0,a)K(t;a) \text{ where } K(t;a) = \exp(\sum_{k=1}^{m}s_k(t)h_k(a));$$

the population size $N(t) = N(0)M_0(\mathbf{s}(t))$,

the values of regulators at $t$ moment $H_k(t) = E^t[h_k] = \partial \ln M_0(\mathbf{s}(t))/\partial s_k$

and the current system distribution
$P(t,a) = P(0,a)K(t;a)/E^0[K(t;.)]$ with $E^0[K(t;.)] = M_0(\mathbf{s}(t))$.

The last formula gives the solution of FHW-equation (3.6). Technically the described approach is useful only if the rank of the fitness matrix $W$ is significantly smaller then its dimension, $m < n$. The approach is especially useful for infinitely dimensional system (3.6). Let us remark that, in general, the fitness matrix can not be known exactly, but its elements can be well approximated by expression (3.7) with small $m$. For example, if $W_{ab} = w_a w_b$ for all $a,b$, then the initial many-dimension (or even infinite-dimension) system (3.6) is reduced to a single ODE. This case corresponds to a well-known example of a population in the Hardy-Weinberg equilibrium.

## 4. Discussion

In this paper we formulate and apply a method that allows us to effectively solve a wide class of replicator equations and corresponding models of selection systems. Most of these models have a form of many (or infinitely) -dimensional systems of differential equations. Some theorems of existence and uniqueness and asymptotic behavior of solutions to particular classes of such equations were established earlier and many particular models were studied, however, to the best of our knowledge, no general methods for solving the RE analytically (except for linear cases) were known.

The suggested algorithm is based on a recently developed theory of inhomogeneous population models and selection systems with distributed parameters [8]. The model behavior may be different and even counter intuitive depending on the initial distribution even for simplest Malthusian and logistic models.

We have applied the method to some replicator equations known from literature, such as fundamental Fisher-Haldane-Wright genetic equation. We hope that this paper may be useful for understanding dynamic peculiarities of solutions of replicator equations and the crucial role of the initial distributions; we also hope that the general method and particular examples presented here can help study replicator equations, which appear in different areas of mathematical biology.